\documentclass[11pt,a4paper]{article}
\usepackage[utf8]{inputenc}
\usepackage{apacite}
\usepackage{authblk}
\usepackage[round]{natbib}
\usepackage{tabularx}
\usepackage{cancel}
\usepackage{multirow}
\usepackage{supertabular}
\usepackage{algorithmic}
\usepackage{algorithm}
\usepackage{amsmath}
\usepackage{float}
\usepackage{rotating}
\usepackage{graphicx,subfigure}
\usepackage[T1]{fontenc}
\usepackage{adjustbox}
\usepackage{url}
\usepackage{array, booktabs, multirow}
\usepackage{threeparttable}

\title{\bf Spectral decomposition and extremes of atmospheric meridional energy transport in the Northern Hemisphere midlatitudes}

\author[1]{Valerio Lembo}
\author[2,3]{Gabriele Messori}
\author[4]{Rune Graversen}
\author[1,5,6]{Valerio Lucarini}

\affil[1]{CEN-Meteorological Institute, University of Hamburg, Hamburg, Germany}
\affil[2]{Department of Earth Sciences, Uppsala University, Uppsala, Sweden}
\affil[3]{Department of Meteorology and Bolin Centre for Climate Research, Stockholm University, Stockholm, Sweden}
\affil[4]{Department of Physics and Technology, University of Troms\o, Troms\o, Norway}
\affil[5]{Department of Mathematics and Statistics, University of Reading, Reading, United Kingdom}
\affil[6]{Centre for the Mathematics of Planet Earth, Department of Mathematics and Statistics, University of Reading, Reading, United Kingdom}

\begin{document}
\maketitle

\begin{abstract}
\small The atmospheric meridional energy transport in the Northern Hemisphere midlatitudes is mainly accomplished by planetary and synoptic waves. A decomposition into wave components highlights the strong seasonal dependence of the transport, with both the total transport and the contributions from planetary and synoptic waves peaking in winter. In both winter and summer months, poleward transport extremes primarily result from a constructive interference between planetary and synoptic motions. The contribution of the mean meridional circulation is close to climatology. Equatorward transport extremes feature a mean meridional equatorward transport in winter, while the planetary and synoptic modes mostly transport energy poleward. In summer, a systematic destructive interference occurs, with planetary modes mostly transporting energy equatorward and synoptic modes again poleward. This underscores that baroclinic conversion dominates regardless of season in the synoptic wave modes, whereas the planetary waves can be either free or forced, depending on the season.
\end{abstract}
 \clearpage

\section{Introduction}
The meridional energy transport from the low to the high latitudes is a key control of both the time mean structure (e.g. \citet{Budyko1969,Sellers1969,Stone1978b,Stone1978a}) and variability (e.g. \citet{BJERKNES19641}) of the Earth's climate. Both the atmosphere and oceans contribute to this transport, with the former accomplishing the larger fraction in the mid to high latitudes (e.g \citet{Trenberth2001}). When considering long-term averages, the horizontal convergence of energy transport performed by the atmosphere and oceans compensates the inhomogeneities of the radiative energy budget at the top of the atmosphere. The very existence of such energy transport defines the climate as a non-equilibrium system, whose fluid motions are forced by the differential absorption of solar radiation between low and high latitudes \citep{Lucarini2014}. Despite the enormous efforts dedicated to improving the skill of state-of-the-art climate models, considerable intra-model discrepancies exist in the description of the meridional energy transport in control runs as well as in climate change scenarios \citep{Lucarini2011b,Lucarini2014,Lembo2018}. \\
The atmospheric energy transport has been traditionally decomposed into a contribution from transient motions, one from stationary waves, and one from the mean meridional circulation (e.g. \citet{Starr1954,Lorenz1967}). Locally, the transient component displays a very large variability, with sporadic extreme episodes exceeding the mean transport values by orders of magnitude \citep{Swanson1997,Messori2013,Messori2014,Messori2017}. Such behaviour underlies a number of important dynamical processes in the atmosphere, from storm track variability \citep{Kaspi2013,Novak2015} to rapid Arctic warming events \citep{Graversen2011,Woods2016,Messori2018}. In this decomposition, the transient component is defined in terms of instantaneous deviations from a long term climatology and effectively includes contributions from all zonal wavelengths (cf. \citet{Starr1954,Peixoto1992}).\\
Recently, an alternative decomposition has been proposed, whereby the meridional energy transport is split into contributions from distinct zonal length-scales (\citet{Graversen2016} and references therein). Interactions between energy transports at different length-scales have been shown to play a crucial role in the energy budget of the high latitudes (e.g.
\citet{Graversen2016,Baggett2016,Goss2016}). Therefore, it is important to draw a clear picture of the role played by the different wavelengths in contributing to the meridional energy transport variability in the extratropics. This approach has strong support from quasi-geostrophic theory: it is well-known that baroclinic instability, forced by meridional gradients of temperature, is active only within a specific range of scales, which define synoptic variability. The theory predicts that the conversion of available potential energy into kinetic energy and the existence of a northward energy transport come hand in hand \citep{Pedlosky1982}. The nature and the energetics behind planetary waves, which are characterized by larger spatial scales and longer time scales, is somewhat less clear. 
Long waves are usually seen as barotropic features fuelled by the inverse turbulent cascades (cfr. \citet{Salmon1980}); indeed, they can alternatively result from baroclinic processes modified and modulated by orography and/or by large scale thermal contrasts \citep{Legras1985,Benzi1986,Speranza1986,Bordi2004}. While these results are extremely relevant, they are mostly confined to linearized theories based on the quasi-geostrophic approximation. It is then crucial to look at atmospheric energy transport at different spatial scales in actual atmospheric data.\\
In this study, we analyze systematically the time-mean and the variability of such scale-dependent transports using the European Centre for Medium-Range Weather Forecasts (ECMWF) ERA-Interim data, with the outlook of applying this approach for inter-comparing climate models and analyzing the impact of climate change on the atmospheric circulation. Our analysis is structured as follows. In Sect. \ref{sec:DataM} we provide a detailed description of the
energy transport decomposition applied here. In Section \ref{sec:Results} we investigate the meridional energy transport at the different spatial scales and underscore the relevance of our approach for better understanding the transport's variability. Section \ref{sec:Results1} addresses the temporal and spatial variability of meridional energy transport as a whole. Section \ref{sec:Results3} is specifically devoted to the seasonal differences in the mean meridional, planetary and synoptic components of the transport along with extreme poleward and equatorward  transport events. We discuss and summarize our findings in Section \ref{sec:Concl}.

\section{Data and Methods}
\label{sec:DataM}
\subsection{Data}
We base our study on 33 years of data from ECMWF's ERA-Interim Reanalysis (Jan 1979-Dec 2012), at a 6h time resolution and T255 horizontal resolution \citep{Dee2011}. The meridional energy transport is integrated vertically over 60 hybrid levels. Due to the non-conservation of mass in reanalysis data \citep{Trenberth1991}, at each time step we apply a barotropic mass-flux correction to the wind field and then compute the energy transport from the updated field \citep{Graversen2006}. The analysis is conducted over a mid-latitude domain spanning the latitudinal belt between 30$^{\circ}$N and 60$^{\circ}$N.

\subsection{The Meridional Energy Transport and its Decomposition}
The energy transport across a given latitude $\phi$ is computed as the product of meridional velocity $v$ with moist static energy (MSE) $H$ (e.g. \citep{Neelin1987})\footnote{Note that its should be more correctly referred to as enthalpy, see for e.g. \citep{Lucarini2011b}.} and kinetic energy:
\begin{equation}
\tau(\phi) = \oint\int_{p_{s}}^{0}vE \frac{dp}{g}dx = \oint\int_{p_{s}}^{0}v(H + \frac{1}{2} \mathbf{v^2})
\frac{dp}{g}dx
\label{transpint}
\end{equation}
where:
\begin{equation}
H = L_{v}q + c_{p}T + gz
\end{equation}
Here $\phi$ is a given latitude, $p$ is pressure, $p_{s}$ is surface pressure, $\mathbf{v}$ is horizontal velocity, $g$ is the gravitational acceleration, $dx$ indicates the differential of space along longitudes, $L_{v}$ is the latent heat of vaporization, $c_{p}$ is the specific heat capacity at constant pressure, $T$ is the absolute temperature, $q$ is the specific humidity, and $z$ is the geopotential height. Our sign convention is therefore that poleward transport in the Northern Hemisphere is positive and equatorward transport is negative. A crucial assumption of this framework is that the hydrostatic approximation is valid.\\ 
In this study we decompose the transport coming from the correlation of the $E$ and $v$ fields into contributions from different zonal wavelengths, following the method proposed by \citep{Graversen2016}. A generic field $\Psi(t,\phi,\lambda)$ can be written as a Fourier series as follows:
\begin{equation}
{\Psi(t,\phi,\lambda)} = \frac{a_0^L}{2}+\sum_{n=1}^{N}\left\lbrace a_n^{\Psi}(t,\phi) \cos\left(\frac{n2\pi \lambda}{d}\right) +
b_n^{\Psi}(t,\phi) \sin\left(\frac{n2\pi \lambda}{d}\right) \right\rbrace,
\label{fourier}
\end{equation}
where $N$ is the maximum resolved zonal wavenumber, and the Fourier coefficients are defined by:
\begin{eqnarray}
a_n^{\Psi}(t,\phi) = \frac{2}{d}\int d\lambda {\Psi}(t,\phi,\lambda) \cos\left(\frac{n2\pi \lambda}{d}\right) \\
b_n^{\Psi}(t,\phi) = \frac{2}{d}\int d\lambda  {\Psi}(t,\phi,\lambda)  \sin\left(\frac{n2\pi \lambda}{d}\right)
\label{fourcoeff}
\end{eqnarray}
with $d = 2\pi R\cos(\phi)$, $R$ the Earth's radius, $\lambda$ and $\phi$ longitude and latitude, respectively. We then obtain the decomposition of the meridional energy transports as a function of zonal wavenumber as:
\begin{eqnarray}
\tilde{\mathcal{F}}_0(\phi) = d \int_0^{p_s} \frac{1}{4} a_0^v a_0^E \frac{dp}{g} \hspace{4.6cm} k=0,\\
\tilde{\mathcal{F}}_k(\phi) = d \int_0^{p_s} \frac{1}{2} \left(a_k^v a_k^E + b_k^v b_k^E \right) \frac{dp}{g}
\hspace{2.3cm} k=1,...,N
\label{transdecomp}
\end{eqnarray}
where superscripts $v$ and $E$ refer to meridional velocity and total energy, respectively, $\tilde{\mathcal{F}}_k$ is the contribution to the meridional heat transport coming from the waves with wavenumber $k$. We partition the zonal wavenumbers $k$ as follows: meridional circulation ($k=0$), planetary waves ($k=1-5$), synoptic waves ($k=6-10$) and higher-order waves ($k=11-20$). Note that higher wavenumber waves are of comparatively little relevance in terms of meridional energy transport. Such a partition is motivated by the fact that we aim at linking the wavenumber approach to the conventional transient-stationary classification.\\
The formulation of the energy transports (Eq. \ref{transpint}) includes transport associated with fluctuations in the net meridional mass transport, which may be large when considering high temporal resolutions (cf. \citet{Liang2018}). These fluctuations are at least the same order of magnitude as the net energy transports. A decomposition of the total transport into a mass-related and a purely energetic component can be performed as follows. First, $vE$ is decomposed into zonal mean and eddy parts as:
\begin{equation}
[vE] = [v][E] + [v^*E^*]
\end{equation}
where [x] is the zonal mean and $x^*$ is the zonal deviation from such mean. It can be further decomposed into time mean and transient components:
\begin{equation}
[\overline{vE}] = [\bar{v}][\bar{E}] + \overline{[v]'[E]'} + [\bar{v^*}\bar{E^*}] + [\overline{v'^*E'^*}]
\end{equation}
where $\bar{x}$ and $x'$ are the time mean and time perturbation, respectively. We can interpret the first term on the right-hand-side as the meridional circulation, the second and the fourth terms as the transient eddies (the former being generally negligible), and the third term as the stationary eddies.
We can further decompose the first term into a vertical mean and a deviation from the vertical mean component:
\begin{equation}
\{[\bar{v}][\bar{E}]\} = \{[\bar{v}]\}\{[\bar{E}]\} +{[\bar{v}]^\dagger[\bar{E}]^\dagger}
\label{meandecomp}
\end{equation}
where $\{x\}$ and $x^\dagger$ indicate the vertical mean and the deviation from the vertical mean, respectively. Of the two r.h.s. terms, only the second one measures the zonal mean meridional energy circulation, the first term being associated with the zonal mean meridional mass flux, which in turn depends on the chosen zero energy reference level. For long term means (i.e. weeks or longer), the first term is negligible, but that is not the case when dealing with 6-hourly values. Rewriting Eq. \ref{meandecomp} for the \textit{instantaneous} zonal mean energy component, we obtain:
\begin{equation}
\{[v]^\dagger[E]^\dagger\} = \{[v][E]\} - \{[v]\}\{[E]\}.
\label{instdecomp}
\end{equation}
In this framework, the zonal mean meridional energy transport is thus obtained by removing from the overall zonal mean transport a term accounting for energy transport associated with the total mass flux, which is dependent on the zero energy reference level. This correction for the mass fluxes is obviously applied only to $k=0$.\\
We define poleward transport extremes as events which fall in the top 2.5 percentiles of the transport distribution for the full domain and time period considered. We further define anomalous equatorward extremes as events falling in the bottom 2.5 percentiles. The choice of threshold for defining the extremes balances the competing requests of having a sufficiently large sample of events and selecting events which are significantly different from median events, so as to justify calling them extremes. Similar results to those described below were also achieved by using a 5\% threshold. Note also that the terms "poleward" and "equatorward" extremes refer to the sign of the anomalies relative to the median value of the distribution, rather than to the sign of the extreme transport itself. However some, but not all, of the anomalous equatorward events correspond to a net equatorward transport. When we observe a net equatorward meridional energy transport, the entropy production due to horizontal heat transports is temporarily negative, as the heat is transported against the temperature gradient (cf. \citep{Lucarini2011a}). Clearly, this is not a violation of the second law of thermodynamics, as the lion's share (about $90\%$) of the entropy production is due to vertical heat exchanges \citep{Lucarini2011a,Lucarini2014}.

\section{Results}
\label{sec:Results}
\subsection{Variability of the Zonally Integrated Meridional Energy Transport}\label{sec:Results1}
\begin{figure}[ht]
\centering
\includegraphics[width=0.85\textwidth]{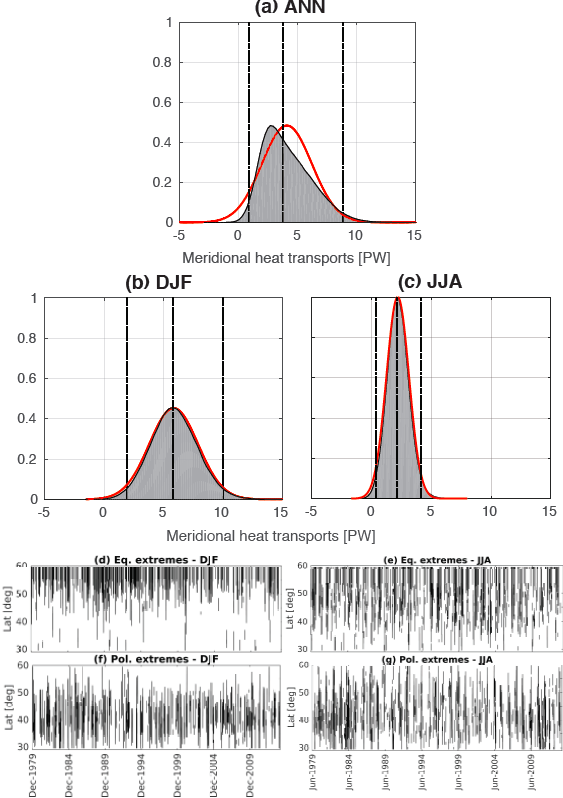}
\caption{\label{fig1} Normalized histograms of 6-hourly annual (ANN) (a), DJF (b) and JJA (c) meridional energy transports in ERA-Interim for the 30$^\circ$--60$^\circ$ N latitudinal channel. The histograms are normalized so that the gray areas are equal in (a)--(c). The normalized Gaussian fits corresponding to the same mean and standard deviation are shown in red. The black vertical dashed lines denote the bottom and top 2.5$^{th}$ percentiles, and the median value of the distribution. Hovm\"oller plots indicating the latitudinal location of extreme events are shown for events (d) below the bottom 2.5$^{th}$ percentile in DJF, (e) below the bottom 2.5$^{th}$ percentile in JJA, (f) above the top 2.5$^{th}$ percentile in DJF, (g) above the top 2.5$^{th}$ percentile in JJA.}
\label{fig:transp}
\end{figure}
The probability density functions (PDFs) of mass-corrected non-decomposed local meridional heat transports for the whole 1979-2012 period and all latitudes in the 30$^{\circ}$N--60$^{\circ}$N band are displayed in Figures \ref{fig1}a--\ref{fig1}c as normalized histograms for the whole year, DJF, and JJA, respectively. The annual PDF (Figure \ref{fig1}a) is clearly different from that of a Gaussian distribution with same mean and standard deviation (red curve). The median is smaller than the Gaussian fitted distribution, while the equatorward tail is thinner and the poleward tail thicker. This skewed distribution is partly the result of different contributions from the winter and summer seasons. In fact, the latter two single-season distributions are both near-Gaussian (Figure \ref{fig1}b, c) but display very different standard deviations and median values. The values of the medians are $5.9$ $PW$ and $2.2$ $PW$ for DJF and JJA, respectively. The bottom and top 2.5$^{th}$ percentiles for the two seasons are $1.9$--$10.0$ $PW$ and $0.4$--$4.1$ $PW$, respectively. In what follows, we focus on the properties of the seasonally stratified energy transports and their extremes.\\
The extreme transports are primarily associated with eddies (cfr. Fig. \ref{fig:transp}b here with Fig. 1c in \citet{Messori2013}). They display a large temporal variability and a strong meridional coherence (Fig. \ref{fig1}d--g). The rapid alternation of positive and negative extremes along the time dimension is immediately evident. While beyond the scope of the present analysis, we note that the transport extremes appear to cluster during periods of one or more months, with gaps in between. This low-frequency modulation of the transport has been seldom discussed (e.g. \citet{Novak2015,Messori2017}). Previous investigations have also noted a spatial clustering of transport extremes \citep{Messori2015}. 

\subsection{Analysis of the Seasonal Cycle and Extremes of the Meridional Energy Transport}\label{sec:Results3}

\begin{figure}[ht]
\centering
\includegraphics[width=\textwidth]{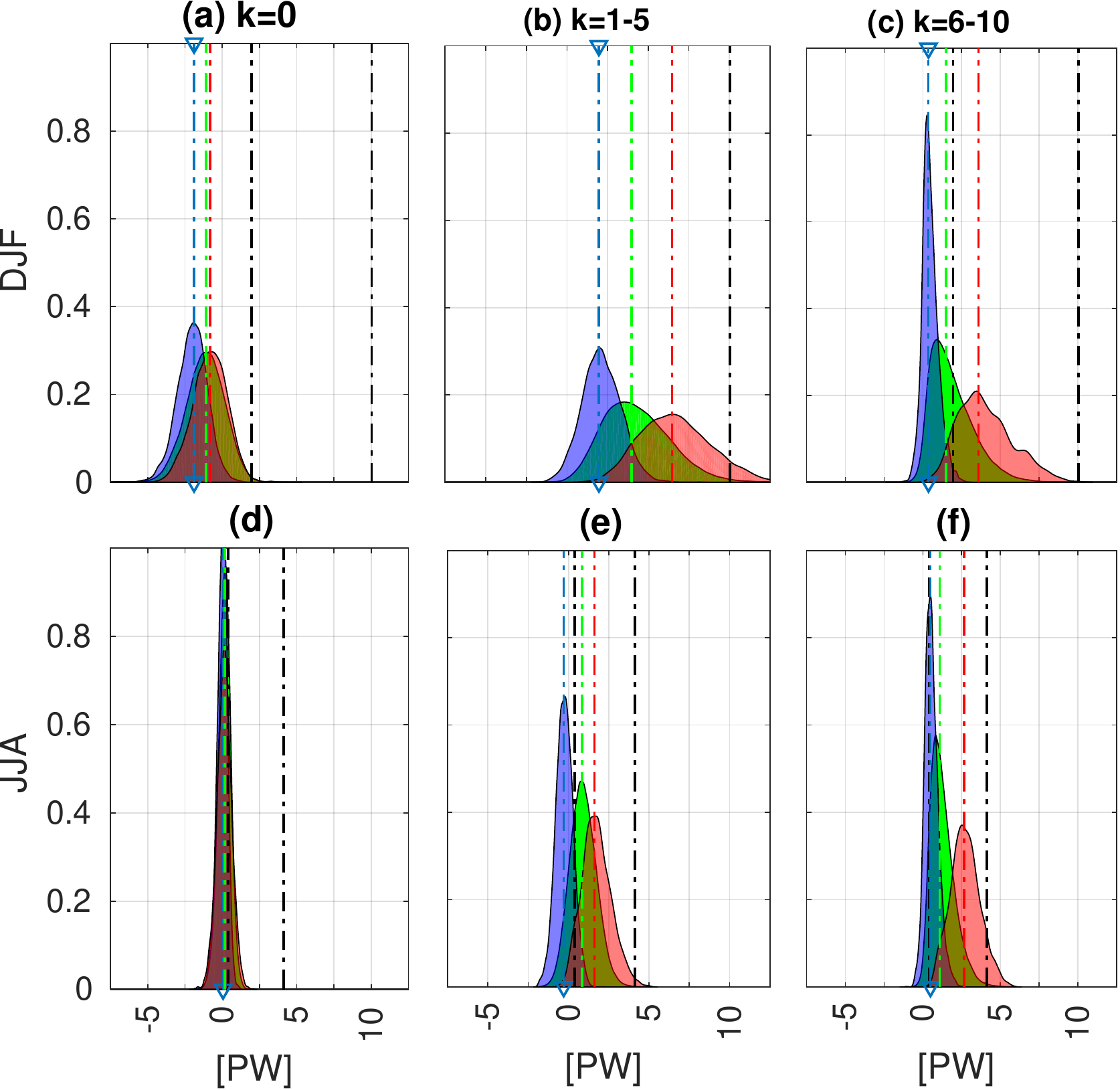}
\caption{\label{fig3}PDFs of meridional energy transports for the 30$^\circ$--60$^\circ$ N latitudinal channel, decomposed by groups of zonal wavenumbers. PDFs of the total transport are in green. The transports corresponding to total transport extremes below the bottom 2.5$^{th}$ and above the top 2.5$^{th}$ percentiles are shown in blue and red, respectively, for DJF (a--c) and JJA (d--f). The wavenumber ranges are: k=0 (a,d), k=1--5 (b,e), k=6--10 (c,f). The dashed vertical lines denote the bottom and top 2.5$^{th}$ percentiles of the total transport (black), the median of the PDFs for the bottom 2.5$^{th}$ percentiles (blue), all events (green) and the top 2.5$^{th}$ percentiles (red). PDFs are estimated via a non-parametric kernel distribution according to a Silvermann test for the modes of a distribution.}
\label{fig:pdfyr}
\end{figure}

We next analyze the contributions of the different scales to the meridional energy transport (see Section \ref{sec:DataM}), separated by season and spatial scale (Figure \ref{fig:pdfyr}). Figure \ref{fig1} shows that the meridional energy transports can achieve very large poleward values, especially in winter. Nevertheless, the zonal mean transport (green PDFs, $k=0$) is on average close to zero during both seasons -- albeit with a preference for equatorward transport in DJF (Figure \ref{fig:pdfyr}). The southward transport by the meridional circulation is consistent with the thermally indirect Ferrel cell prevailing at these latitudes. Without the mass correction that we here apply, the 6-hourly PDF of the zonal mean component would span a similar range as the eddy components, while retaining a near-zero median value (not shown). The total transports contributed by the planetary and synoptic waves (green PDFs, k=1--5 and k=6--10, respectively) are shifted towards positive values. This is especially evident during DJF, in line with the seasonal cycle of the total transport (e.g. \citet{Fasullo2008}) and is the result of the larger availability of potential energy for baroclinic conversion in the winter season (e.g. \citet{Grotjahn2015}). The planetary and synoptic wavelengths thus account for the bulk of the poleward energy transport, and rarely produce equatorward transports. This is associated with a positive skewness of the PDFs, peaking in DJF. The rare cases where equatorward transports do occur are extremely relevant from the point of view of the thermodynamics of the atmosphere because they are associated with weak or counter-gradient eddy transports, as discussed later in this section.\\
A comparison with the meridional transports in the 30S-60S (cfr. Text S1) suggests that the weakening of the synoptic and planetary-scale transports in the warm season is primarily a feature of the NH, whereas the seasonality is much less pronounced in the SH. In the latter, the planetary eddies play a large role year-round. At first glance, the relevance of the planetary eddy contribution might be surprising, given that previous findings point to weaker stationary waves in the planetary-scale range for the SH, e.g. \citep{DellAquila2007}. Hence, our results suggest that the contribution by planetary waves to the meridional heat transport is dominated by travelling waves in the SH (whereas we expect orographically-forced planetary waves in the NH).\\
We next focus on the extreme values of the total transport, as defined in Section \ref{sec:DataM}. In both DJF and JJA, some equatorward extremes (blue PDFs) in the total transport correspond to \textit{anomalous} conditions where the net transport is negative and hence equatorward. These extremes occur more frequently in summertime (cfr. Figures \ref{fig1}b and \ref{fig1}c). The mean meridional circulation displays both positive and negative contributions to the transport -- for both positive and negative extremes -- albeit in DJF with the distributions shifted towards the sign of the extremes being considered. However, regardless of the season, there is a very large overlap between the PDFs of opposing extremes. This means that it is not unusual for the transport performed by the mean meridional circulation during some poleward extremes (red PDFs in Figure \ref{fig:pdfyr}) to be smaller than that performed during some equatorward extremes. The overlap between the PDFs associated with the two extremal tails of the total transport PDF is much smaller for the planetary waves. Even for equatorward extremes, the net transport performed by the planetary waves is almost always poleward, especially in DJF. Similar considerations hold for the synoptic waves. The PDFs further highlight that, even in the dynamically steadier summer season, there can be extremely strong planetary to synoptic scale events, effecting large instantaneous transports. Thus, the emerging picture is that the contribution from the planetary and synoptic scales is the main discriminant between the poleward and equatorward extremes, while the mean meridional circulation can contribute with changes of either sign to either class of extremes.  As mentioned in Text S2, the PDFs are relatively similar in shape when specific latitudes are considered (45N, 30N and 60N in the SI). Nevertheless, both the planetary and synoptic components are weaker at 45N, whereas the synoptic and planetary components are respectively stronger and weaker at 30N (and vice versa at 60N). Figure \ref{fig1}d suggests that the equatorward extremes might play a relevant role also poleward of 60N, at least in winter. However, an investigation of latitudes poleward of 60N (not shown) suggests that the relative importance of planetary vs. synoptic eddy transports is similar to what found for 60N.\\

\begin{figure}[ht]
\centering
\includegraphics[width=\textwidth]{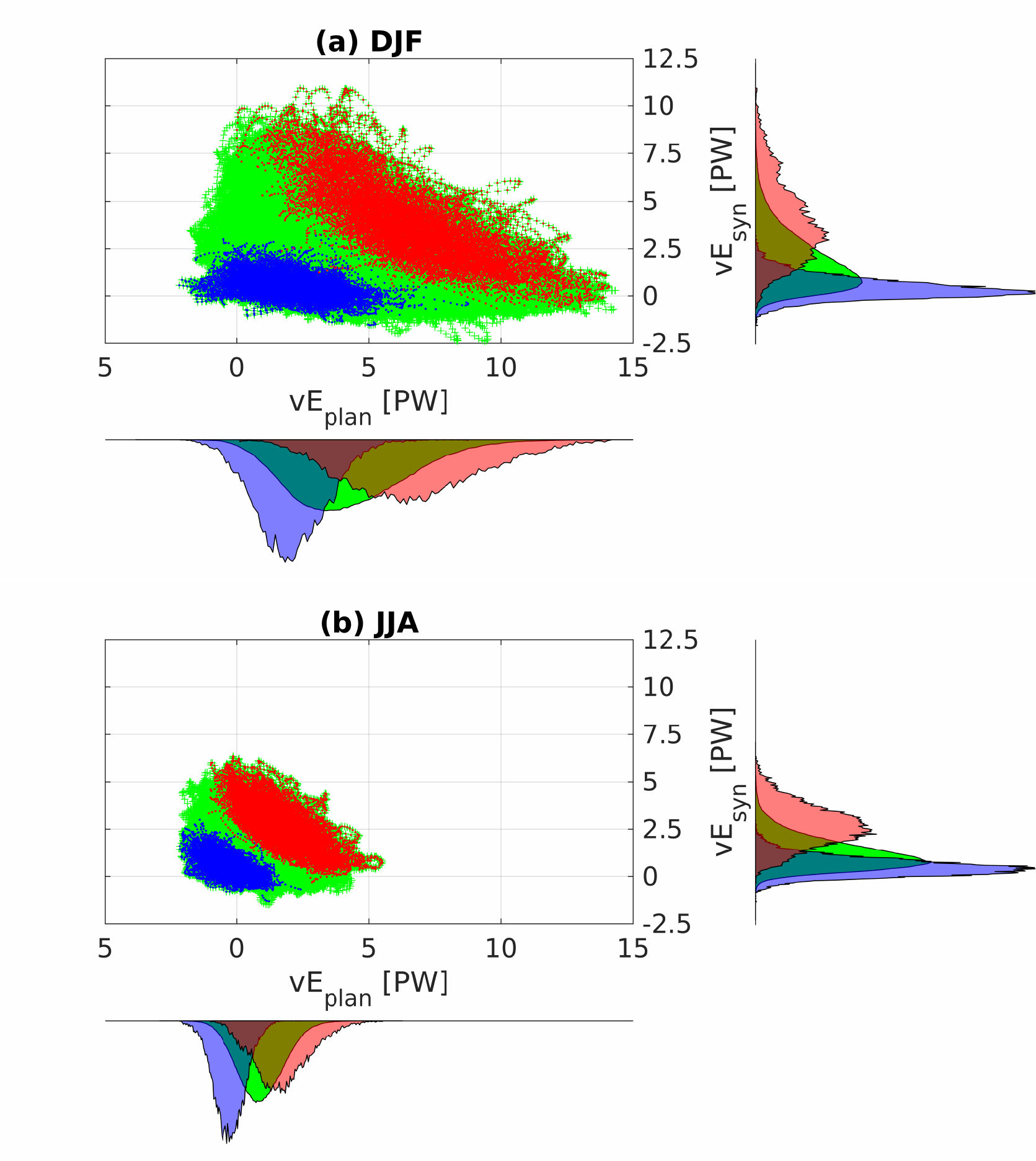}
\caption{Scatter plots of synoptic vs. planetary scale meridional energy transports for (a) DJF and (b) JJA. All events (green crosses), and those below the bottom 2.5$^{th}$ (blue dots) and above the top 2.5$^{th}$ (red dots) percentiles are shown. Next to each axis, the PDFs of the respective components (as in Figure \ref{fig:pdfyr}) are also shown.}
\label{fig7}
\end{figure}
\begin{figure}[ht]
\includegraphics[height=0.8\textheight]{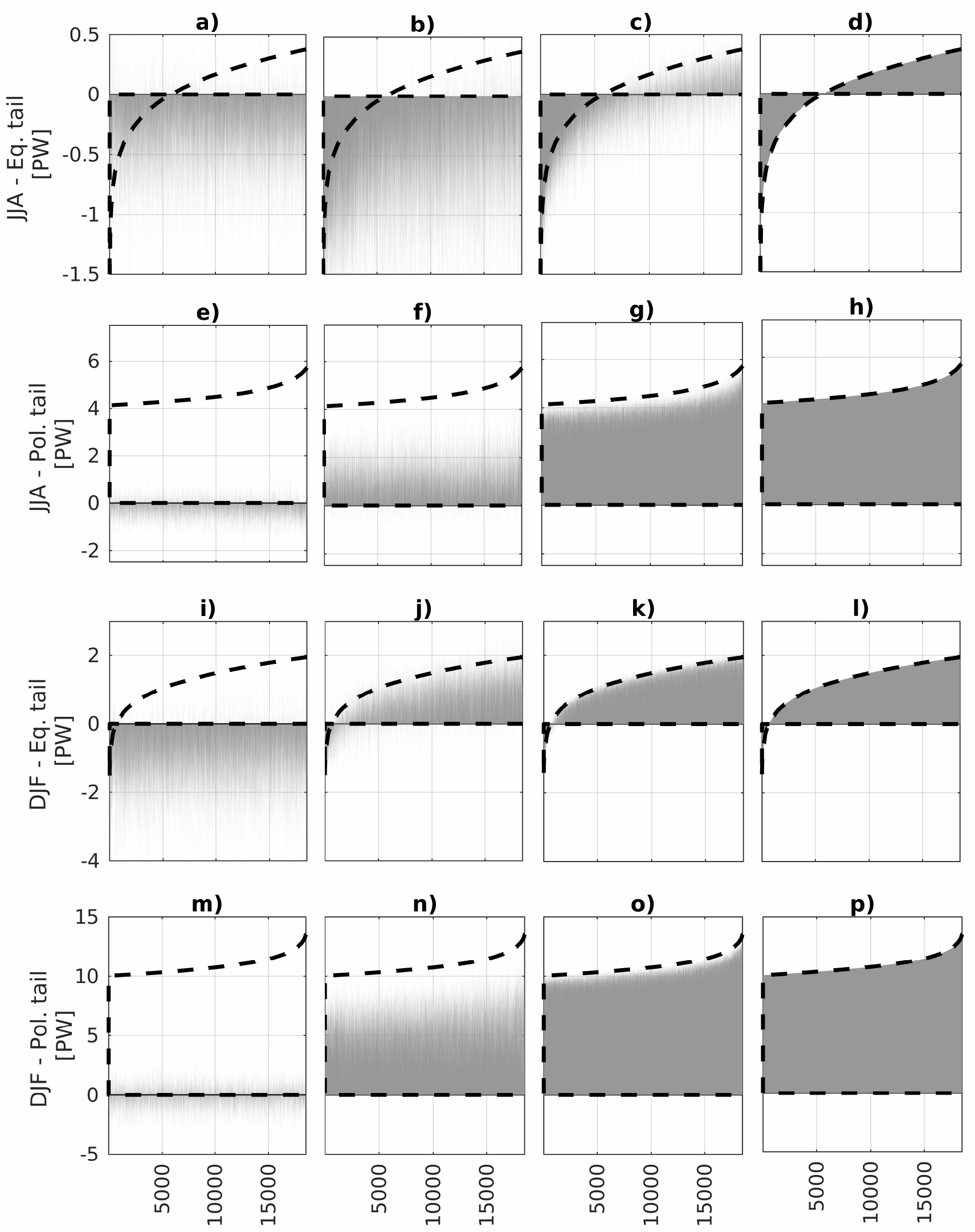}
\caption{Bottom and top 2.5$^{th}$ percentile extreme events in the 30N-60N latitudinal channel latitude for JJA (a--h) and DJF (i--p). The events are sorted based on the total transport from the largest to the smallest one, for k=0 (zonal mean; panels a,e,i,m), k=0--5 (including planetary waves; panels b,f,j,m), k=0--10 (including synoptic waves; panels c,g,k,o), k=0--20 (including all waves; panels d,h,l,p). The dashed lines encompass the values of the total meridional transports. Values are in PW.}
\label{fig6}
\end{figure}

The above analysis motivates further investigating the interplay between simultaneous planetary and synoptic transport components. As shown in Figure \ref{fig7}, when extreme transports occur there is an inverse relationship between the two: stronger planetary transports correspond to weaker synoptic transports and vice versa. Further, there is a systematic shift towards higher values for the poleward and lower values for the equatorward extremes, and a larger variability in the poleward than in the equatorward extremes is found in both the planetary and synoptic wave components. This highlights that while neither equatorward nor poleward transport extremes at planetary and synoptic scales typically co-occur, the extremes are the result of a systematic shift of the conditional probability distributions of the transports. It can be argued that some kind of competition occurs between planetary and synoptic extremes, so that the overall transport can be effected by an extreme synoptic transport, an extreme planetary transport, or co-occurring intermediate values of the two. However, while in the case of equatorward extremes the two can be of opposite sign, and thus interfere destructively, poleward extremes are characterized by a predominantly constructive interference. There is a clear difference between DJF and JJA (cfr. Figure \ref{fig7}), with the overall synoptic transports (green PDFs on the ordinates) being relatively independent of the season, and the planetary transports (green PDFs on the abscissa) contracting significantly in JJA when compared to DJF. This points to a substantially different mechanism of APE conversion in the two seasons, as we discuss further below.\\
The contribution of the different scales to the transport extremes is exemplified in Figure \ref{fig6}, where the transports in the poleward and equatorward tails of the distribution are sorted according to the value of the non-decomposed transport. In both DJF and JJA, poleward extremes are characterized by weak zonal mean transports of both signs (panels e,m). The extremes are almost entirely determined by the planetary and synoptic-scale transports. The former are most relevant in DJF (panels n,o), while the latter dominate in JJA (panels f,g). This is reflected in the planetary transports shown in Figure \ref{fig7}. We hypothesize that this seasonality may partly correspond to the seasonality of the blocking events, whose frequency within our domain is strongly reduced in JJA. It has also been found that the weak meridional temperature gradient in JJA concentrates baroclinic conversion in a comparatively narrow spectral range centered around the synoptic scales \citep{Holton2004}. Looking at the equatorward extremes, the zonal mean component is almost always negative (equatorward), regardless of season (panels a,i), and reach values in excess of -4 $PW$. This is up to two orders of magnitude larger than the average contribution by the zonal mean component, amounting to about 10$^{-1}$ $PW$. Adding the planetary component shifts most values to positive in DJF (panel j), while it strengthens the equatorward values in JJA (panel b). This partially contrasts with the pattern emerging from the analysis of the SH, where the planetary eddies oppose the zonal mean transport in the equatorward extremes of the warm season (Text S1). The synoptic scales further increase the poleward transport during DJF (panel k), while higher wavenumbers contribute little to the picture. In JJA, on the other hand, the synoptic scales are instrumental in shifting the net transport of many extremes from negative to positive (panel c), and the higher wavenumbers provide a significant contribution to this (panel d). As already remarked above, in DJF the planetary component is therefore more relevant than the synoptic one. In JJA, planetary waves instead appear to be passive, in the sense that they are not able to convert available potential energy into kinetic energy by acting counter-gradient.

\section{Discussion and Conclusions}
\label{sec:Concl}
In this study we present a wavenumber decomposition of the vertically integrated meridional atmospheric energy transport in the Northern Hemisphere extratropics, in the spirit of the spatio-temporal decomposition of the atmospheric variability (cfr. \citet{Hayashi1979,Ulbrich1991,DellAquila2005,Graversen2016}). The decomposition performed here is more informative than the traditional separation into transient flow patterns, stationary waves, and the mean meridional circulation (e.g. \citet{Starr1954}), where each component effectively includes contributions from all spatial scales. This made it difficult to separate the role of the different atmospheric motions in driving the transport's variability and extremes. As it is well known from quasi-geostrophic theory, the mechanisms responsible for the meridional transport of energy act preferentially on specific spatial (and temporal) scales. The strength of our approach is that it enables us to separate the roles of individual wavenumbers, which in the analysis are grouped into: meridional circulation (k=0), planetary waves (k=1--5), and synoptic waves (k=6--10). We also focus on poleward and equatorward extremes, defined here as the top and bottom 2.5 percentiles of the total transport distribution, respectively.\\
The different wavenumbers reflect the seasonal cycle of the net transport, whose strength and variability in the boreal hemisphere peak in the corresponding winter season (whereas the seasonality is less pronounced in the austral hemisphere). In the JJA season the poleward transport performed by planetary waves is relatively weak, as a result of the fact that the reduced baroclinicity narrows the range of scales where baroclinic conversion can take place. This is particularly evident when picking some specific latitude in the centre of the 30-60N latitude channel (cfr. Text S2).\\
For poleward extreme events, we observe that a compensation mechanism acts between synoptic and planetary waves: while the majority of poleward extremes display above average planetary and synoptic-scale transports, their magnitudes are inversely correlated. However, both are predominantly positive so that their interference is constructive. The mean meridional circulation generally plays a marginal role, and is on average weakly equatorward. Instances of constructive interference between the poleward transports driven by transient and stationary waves have been previously associated with heat and moisture transport to the high latitudes \citep{Goss2016}, as well as more meridionally oriented circulation regimes that divert synoptic systems poleward \citep{Baggett2016} and large values of zonally integrated poleward energy transport \citep{Messori2017}. In this context, the decomposition presented here elucidates the contributions of the different groups of wavenumbers. \\
Equatorward extremes follow a different pattern. In winter, the transport performed by the mean meridional circulation is overwhelmingly equatorward, while the higher wavenumbers mostly effect a poleward transport. During summer these extremes correspond to a systematic destructive interference between a mostly equatorward transport by planetary waves and a mostly poleward transport by synoptic and smaller-scale motions. This highlights the predominant role of synoptic eddies in baroclinic conversion. \\
A limit of the present work is in the lack of a detailed investigation of the mechanisms underlying the different flavours of the meridional heat transports. Future studies should investigate the links between the multi-scale interactions highlighted here and dynamical features such as the meandering of the jet stream and the spatial patterns and intensity of blocking events. In addition, an analysis of the thermodynamic features associated with the alternation of constructive and destructive interference could yield interesting insights into the mechanisms driving upscale energy transport from the high to the low latitudes. This could be achieved by performing an analysis of the Lorenz energy cycle scale by scale or, in other words, wavenumber by wavenumber, as done in \citep{Zappa2011} or (using the traditional stationary-transient decomposition) \citep{Lembo2019}. Another line of research would involve verifying how well the latest generation of high-resolution climate models are able to reproduce the scale dynamics underlying the poleward and equatorward transport extremes and whether they are able to provide a good representation of the physical mechanisms behind them. This is key for assessing their ability to simulate the meridional energy transport in different climatic conditions, which is a nontrivial task \citep{Lucarini2011b,Lucarini2014}, and for achieving seamless prediction across timescales \cite{Palmer2008}. Finally, an important outcome of the present study, that we plan to exploit in future research, is that it points to a non-trivial relation between the mechanism of baroclinic adjustment \citep{Stone1978b}, acting scale by scale, and the changes in the meridional temperature field.

\section{Acknowledgments}
Valerio Lembo was supported by the Collaborative Research Centre TRR181 "Energy Transfers in Atmosphere and Ocean" funded by the Deutsche Forschungsgemeinschaft (DFG, German Research Foundation), project No. 274762653. Gabriele Messori was partly supported by a grant from the Department of Meteorology of Stockholm University and by Swedish Research Council grant No. 2016-03724. Rune G. Graversen was partly supported by the Research Council of Norway, project No. 280727. Valerio Lucarini was partially supported by the SFB/Transregio TRR181 project and by the EU Horizon 2020 Blue-Action Project (Grant No. 727852). Data was partly processed at the Stallo supercomputer at the University of Troms\o{} (UiT) provided by the Norwegian Metacenter for Computational Science (NOTUR), projects No. NN9348k and NS9063k. All data has been obtained from the publicly accessible ERA-Interim repository: \url{https://apps.ecmwf.int/datasets/data/interim-full-daily/}.

\end{document}